\documentclass[aps,pre,twocolumn,groupedaddress,showpacs,showkeys]{revtex4-1}
\usepackage[dvips]{graphicx}
\usepackage{cmap}
\usepackage{epstopdf}
\begin{document}

%\preprint{}

\title{Isothermal evaporation rate of deposited liquid aerosols and the SARS-CoV-2 
coronavirus survival}


\author{P.S. Grinchuk, E.I. Fisenko, S.P. Fisenko, S.M. Danilova-Tretiak}
\email[P.S. Grinchuk $\:$]{gps@hmti.ac.by}
\homepage[]{www.itmo.by} 
%\thanks{}
\affiliation{A.V.Luikov Heat and Mass Transfer Institute, National Academy of Sciences of
Belarus,\\ 15 P. Brovka Str., Minsk 220072, Belarus.}


\date{\today}

\begin{abstract}
It is shown that the evaporation rate of a liquid sample containing the culture of coronavirus affects its survival on a substrate. Possible mechanisms of such influence can be due to the appearance of large, about 140 bar, non comprehensive capillary pressures and the associated dynamic forces during the movement of the evaporation front in a sample with the virus. A simulation of isothermal evaporation of a thin liquid sample based on the Stefan problem was performed. The comparison of simulation data and recent experiments on the coronavirus survival on various surfaces showed that the rate of isothermal evaporation of aqueous samples, which is higher for heat-conducting materials, correlates well with the lifetime of the coronavirus on these surfaces.
\end{abstract}

%\keywords{}

\maketitle

%%%%%%%%%%%%%%%%%%%%%%%%%%%%%%%%%%%%%%%%%%%%%%%%%%%%%%%%%%%%%%%%%%%%%%%%%
%\section{Introduction}
%1
In a recent work \cite{COVID-1_Surf}, experimental medical data on the survival of coronavirus on various surfaces have been presented. 
Deposition of liquid samples consisting of water with the coronavirus culture on a surface of various materials is common in everyday life and medicine (plastic, cardboard, stainless steel). 
Surprisingly the lifetime of the virus depends significantly on the substrate material and can differ by almost an order of magnitude. 
For brevity, below we will call these samples aqueous ones, although the composition of the liquid fraction is more complex. 
The mechanisms of influence of a substrate material are not known now. Therefore, in our opinion, it is impossible to explain such behavior only by the specific biological interaction of viruses and materials. 
All experiments \cite{COVID-1_Surf} were carried out in laboratory conditions at a constant temperature and, apparently, constant illumination. 
We suppose that the evaporation of aqueous samples plays an important role in the survival of the coronavirus under given conditions.

%2
The influence of such factors as temperature and humidity on the distribution of coronavirus is undeniable. The highest number of incidents of SARS-CoV-2 coronavirus is observed in countries located from 30 to 50 degrees north latitude \cite{Statistics_COVID}. 
Obviously, these countries are primarily linked by the climatic  
factor in the sense of moderate temperatures and high humidity in winter and spring. Therefore the evaporation rate of water samples with virus is quite low at such conditions.

%3
The resistance of viruses to adverse environmental factors is determined by their structure. 
In biology, it is customary to distinguish between simple and complex viruses. Simple, or non-enveloped, viruses consist of nucleic acid and protein coat (capsid). 
Complex or enveloped viruses outside the capsid are also surrounded by a lipoprotein envelope (supercapsid) outside the capsid, which makes them more vulnerable to adverse environmental factors \cite{Ref3}. 
The coronavirus SARS-CoV-2 is complexly organized, has a lipoprotein membrane, and its size reaches 80 - 160 nm \cite{Ref3,Ref4}. 
The characteristic size of the coronavirus protein whiskers is only 10-20 nm \cite{Ref5}. 
Note that it has already been established that the coronavirus protective membrane is well wetted \mbox{by water \cite{Ref5}.} 

%4-6
It is known that both absolute (AH) and relative humidity (RH) affect various types of viruses \cite{Humidity,Ref7}. 
Currently a number of biological and chemical mechanisms of the influence of composition of aqueous samples on viruses are considered: water activity, surface inactivation, salt toxicity etc. \cite{Humidity}. 
We would like to discuss a purely physical mechanism of influence here. %???
We propose a new hypothesis about the influence of evaporation of aqueous samples with coronaviruses on the survival of coronaviruses. 
We suppose that the movement of the evaporation front through the liquid layer with the virus leads to the cessation of its activity. The resulting local nanoscale curvatures of the liquid surface lead to the appearance of huge local gradients in hydrostatic pressure, which cause damage to the virus. 
The high resistance of viruses to external comprehensive mechanical loads is well known \cite{Ref8,Ref8a}. 
We assumed that the lifetime of the virus during the movement of the evaporation front is significantly reduced due to additional mechanical stresses on some whiskers in a nanolayer of liquid above the virus. 
Indeed, when the evaporation front moves under the action of capillary forces (Laplace pressure), a pressure drop of about 10 bar at a diameter of 80 nm acts on the coronavirus. 
At the same time, a pressure of out 140 bar is already acting on his "whiskers", which have characteristic dimensions of 10-20 nm. 
Inhomogeneous dynamic capillary pressure forces can cause mechanical damage to the virus.  
Observed effect of the decrease in the concentration of viable viruses in aqueous sample during evaporation from a solid substrate in experiments confirms our assumption about existence of the physical mechanism \cite{COVID-1_Surf}.

%7
There is at least a qualitative analogy, confirmed by different experimental data. Significant deformations of a graphene oxide sheet were experimentally discovered in an evaporating micron droplet of water \cite{Ref9_Graphene1}. Graphene is a very strong material, the elastic modulus of which reaches 1 TPa \cite{Ref10_Graphene2}.

In any case, the presence of a virus near the evaporation front violates the symmetry of the evaporation front, which should lead to nanoscale deformations.
At a higher evaporation rate, for example, for a copper substrate, deformations near the evaporation front should increase in amplitude and, as experimental data show \cite{COVID-1_Surf}, individual mechanical properties of each virus type begin to reveal. 
At relatively slow evaporation rates, the difference between kinds of viruses disappears (experiments with aerosols on plastic and stainless-steel substrates with SARS-CoV-2 and SARS-CoV-1).

%8
The purpose of this work is to show, at least semi-qualitatively, the impact of the thermal conductivity of a substrate material on the evaporation of a liquid sample with virus culture. 

%9
Thus, we connect the problem of the virus survival with the problem of evaporation of a liquid sample as we consider the conditions of experiments \cite{COVID-1_Surf}. 
The experiments were carried out at room temperature \mbox{21-23 $^{\circ} C$}. 
The relative humidity during the preparation of the liquid substance with the virus culture was 40\%. 
The survival rate of the coronavirus culture was evaluated for polypropylene, stainless steel from AISI 304 alloy, copper and cardboard. 
Thus various domestic and hospital situations were experimentally modeled. 
Note that the pieces of materials were of arbitrary sizes and thicknesses. The initial concentration of the virus in the biomaterial was $10^{3.4}-10^{3.7}$ or 2500-5000 $ml^{-1}$.  
This corresponded to the typical concentrations observed in the upper and lower respiratory tract of a human. 
Next, a liquid biomaterial with a viral culture in a volume of $V_{d} = 50 \: \mu l$  was applied to the surfaces described above. 
At certain time intervals, smears were taken from the surfaces and the virus concentration was analyzed. 
The virus was considered deactivated at that moment in time when its concentration decreased to a value of units per milliliter. 
Thus, it was experimentally measured that the lifetime of the virus  was  approximately 50 hours on a plastic  substrate, was 24 hours  on  a cardboard substrate, was 30 hours on a) stainless steel substrate, and was 5 hours on a copper substrate \cite{COVID-1_Surf}. 
These results are of great practical importance for epidemiologists in order to reach the effective prevention of the spread of coronavirus \cite{Vaccin_PRE}.

%10
Our experience in solving evaporation problems indicates that the evaporation rate of thin films and ensembles of aerosol particles depends on the thermal conductivity of the substrate \cite{Ref11_Model}. 
The evaporation proceeds faster from more heat-conducting materials and from thinner substrates. 
However, before quantifying and analyzing the experimental data, two points must be made. 
For the objectivity of the results we want to exclude copper and cardboard. 
It is known that copper and its compounds possess both antibacterial and antiviral activity \cite{Ref12,Ref13}. 
At the same time, such activity is absent in stainless steel \cite{Ref13}. 
The highest death rate of the virus was observed on a copper substrate, which leads to an assumption, that it can be the result of the occurrence of a cumulative effect, in which the antiviral properties of the material work in conjunction with the effect of liquid evaporation discussed in this paper. 
It is important to note that the evaporation of a liquid sample occurs most quickly on copper.

%11
Regarding a cardboard, it should be noted that this is a porous material on which, in addition to evaporation, moisture is absorbed and diffused into the material. 
Therefore, the lifetime of a liquid sample with viruses on cardboard is reduced, as well as on any other porous material that absorbs moisture. This qualitative explanation correlates with the well-known fact that respiratory viruses live less on porous surfaces \cite{Ref14_Porous}. 
Therefore, only simulation results for stainless steel substrate and plastic one can be correctly compared.

\begin{figure}
\includegraphics[width=3.2 in,angle=0]{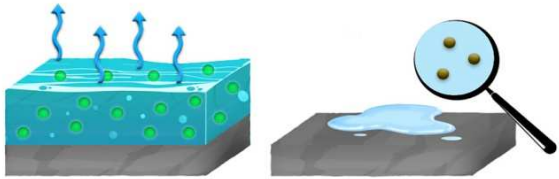}
\caption{The sketch for evaporation of a liquid sample with a viral culture from the solid surface. 
\label{Fig1}}
\end{figure}

%12
We calculate the time of evaporation of the deposited liquid layer of thickness $h$ on a substrate of thickness $H$, taking into account the thermophysical properties of all the materials involved in the process (Fig.\ref{Fig1}). 
This problem is governed by the one-dimensional non-stationary moving boundary problem known as the Stefan problem. 
Brownian diffusion of coronaviruses maintains their approximately uniform distribution in the sample volume, so the use of a one-dimensional approach seems reasonable. 
The temperature $T$ is governed by the heat conduction equation for substrate

%%%%%%%%%%%%%%%%%%%%%%%%%%%%%%%%%%%%%%%%%%%%%%%%%%%%%%%%%%%%%%%%%%%%%%%%%
\begin{equation}
\rho_{s} c_{s} \frac{\partial T_s}{\partial t} = \lambda_{s} \frac{\partial^2 T_S}{\partial z^2} ,
\label{eq1}
\end{equation}
%%%%%%%%%%%%%%%%%%%%%%%%%%%%%%%%%%%%%%%%%%%%%%%%%%%%%%%%%%%%%%%%%%%%%%%%%

\noindent and for liquid layer 

%%%%%%%%%%%%%%%%%%%%%%%%%%%%%%%%%%%%%%%%%%%%%%%%%%%%%%%%%%%%%%%%%%%%%%%%%
\begin{equation}
\rho_{w} c_{w} \frac{\partial T_w}{\partial t} = \lambda_{s} \frac{\partial^2 T_w}{\partial z^2} ,
\label{eq2}
\end{equation}
%%%%%%%%%%%%%%%%%%%%%%%%%%%%%%%%%%%%%%%%%%%%%%%%%%%%%%%%%%%%%%%%%%%%%%%%%

\noindent subject to the isothermal initial conditions 
$\left. T_s(z) \right |_{t=0} = T_0$,
$\left. T_w(z) \right |_{t=0} = T_0$,
and the second kind boundary condition at the boundary of the substrate and the liquid 

%%%%%%%%%%%%%%%%%%%%%%%%%%%%%%%%%%%%%%%%%%%%%%%%%%%%%%%%%%%%%%%%%%%%%%%%%
\begin{equation}
\left. \lambda_{s} \frac{\partial T_s}{\partial z} \right |_{z=H}=
\left. \lambda_{w} \frac{\partial T_w}{\partial z} \right |_{z=H}.
\label{eq3}
\end{equation}
%%%%%%%%%%%%%%%%%%%%%%%%%%%%%%%%%%%%%%%%%%%%%%%%%%%%%%%%%%%%%%%%%%%%%%%%%

The location of the moving evaporating surface of the liquid is governed by the heat balance equation known as Stefan condition

\begin{equation}
\rho_w \left[ c_w T_w(h(t)) + U_w \right] \frac{dh}{dt} =
- \lambda_{w} \frac{\partial T_w}{\partial z} + A_w+A_a.
\label{eq4}
\end{equation}

%13
\noindent Here  $\lambda_{s}$, $\lambda_{w}$ are the thermal conductivity coefficients of the substrate material and the evaporated liquid;  
$U_w$ is the specific heat of vaporization of the liquid; $\rho_s$, $\rho_w$  are the mass density of the substrate and liquid;   
$A_w$ and $A_a$  are the energy flows carried by molecules of an evaporating liquid and a ballast gas, respectively. 
We suppose that evaporation is isothermal, which is typical for the experiments described in \cite{COVID-1_Surf}.
% % % % Figure 1
% % % % % % % % % % % % % % % % % % % % % % % % % % % % % % % % % % % % % %
%\begin{turnpage}
%\end{turnpage}
% % % % % % % % % % % % % % % % % % % % % % % % % % % % % % % % % % % % % % 

%14
Using the approach described in \cite{Ref11_Model}, from the solution of the Stefan problem the following expression can be obtained for the time  $\tau_{ev}$  of the complete evaporation of the liquid sample

\begin{equation}
\tau_{ev} = \frac{\rho_w U_w \: h}{2 (T-T_{lim})} 
\left( \frac{h}{\lambda_w} + \frac{H}{\lambda_s}\right).
\label{eq5}
\end{equation}

%15
\noindent Here $T_{lim}$ is the temperature of the wet thermometer, which depends only on the temperature and humidity near the substrate. 
For experimental conditions in \cite{COVID-1_Surf} the relative humidity is about of 40\% and the difference $T- T_{lim}$ is equal to $14^{\circ} C$. 
It is this temperature difference that provides thermodynamic force for the liquid evaporation. 
It is worthy to emphasize that the temperature of the evaporation surface for quasi steady regime is  approximately equal to $T_{lim}$.

%16
For  further  calculations, we need to determine the thickness of the substrate and the evaporating film one. In [1], unfortunately, the thicknesses of the materials on which the viral substance was applied were not indicated. Therefore, from general considerations about the thicknesses of materials used in everyday life, we will take the thickness of the substrate equal to H = 1 mm. The characteristic thickness of the liquid layer can be estimated only on the basis of the

For further calculations, we need to determine the thickness of the substrate and the evaporating film one. 
In \cite{COVID-1_Surf}, unfortunately, the thicknesses of the materials on which the viral substance was applied were not indicated. 
Therefore, from general considerations about the thicknesses of materials used in everyday life, we will take the thickness of the substrate equal to $H = 1 mm$. 
This is, for example, the thickness of a credit card.
The characteristic thickness of the liquid layer can be estimated only on the basis of the volume indicated in the article as $h \approx V_d^{1/3} = 3.7 \: mm$. 
We assume that the evaporating liquid is close to some properties of water, 
for which  $l_w = 0.60$ W/(m K), $U_w = 2.26$ MJ/kg. 
Then for stainless steel substrate ($\lambda_s =16.2$ W/(m K)) and polypropylene one ($\lambda_s = 0.22$ W/(m K)), the evaporation time of a liquid layer with a thickness of 3.7 mm on plastic is equal to 3190 seconds, and on stainless steel is about 1850 seconds. 
In experiments, the lifetime of the virus on a plastic surface exceeds the lifetime on stainless steel substrate by approximately 1.7 times. 
Calculated the ratio of evaporation times on different substrates is also 1.7 times. 
This proportion depends only on the physical properties of materials and their thicknesses. 
The absolute value of the evaporation time in this case also depends on RH.
Evaporation time increases significantly with increasing RH of the surrounding space, see Exp. (\ref{eq5}). 
No doubts that experiments with dangerous viral substances were carried out in closed volumes during several hours. 
Thus, liquid evaporation increases humidity in the experimental box. 
In our case an increase in evaporation time by an order of magnitude can occur if the relative humidity rises from 40 to 85\%.

%17
The primary experimental data from \cite{COVID-1_Surf} gives the dynamics of the change in the concentration of the virus over time. 
If our assumption is correct, then the death rate of viruses should strongly correlate with the evaporation rate of the liquid sample. 
Below we present the results of quantitative estimates that follow from this assumption. Then we compare them with experimental data \cite{COVID-1_Surf}.

%18
From the solution of the Stefan problem, we can obtain the functional dependence of the height of the evaporating layer versus time for a substrate with poor thermal conductivity

\begin{equation}
h= h_0 - \frac{\lambda_s}{\rho_w U_w } \frac{(T-T_{lim})}{H} t = h_0-\beta \: t,
\label{eq6}
\end{equation}

\noindent and for a substrate with high thermal conductivity

\begin{equation}
h= \sqrt{h_0^2-\frac{\lambda_w}{\rho_w U_w } (T-T_{lim}) t}  = \sqrt{h_0^2-\alpha \: t}.
\label{eq7}
\end{equation}

\noindent Here $h_0$ is the initial layer height on the substrate.

%19
For estimation of the dynamics in the concentration of viruses that are outside a living organism, we use the simplest Malthus model \cite{Ref15_Malthus}.
According to this model, the evolution of the concentration $C$ of viruses is described by the equation

\begin{equation}
\frac{d C}{d t} = k \: C.
\label{eq8}
\end{equation}

\noindent Here $k$ is a parameter characterizing the dynamics of a population change. In the framework of the proposed approach, we believe that the parameter $k$ for the virus population will be proportional to the rate of evaporation $k=k^{\prime} V_{ev}= k^{\prime} \frac{dh}{dt}$. 
Then, for the two limiting cases of evaporation of a viral liquid sample, we have the following equations for the dynamics of the virus population.
The equation for low heat conductive substrate $(\lambda_s < \lambda_w )$

\begin{equation}
\frac{d C}{d t} = - k^{\prime} \frac{\lambda_s}{\rho_w U_w } \frac{(T-T_{lim})}{H} \: C,
\label{eq9}
\end{equation}

\noindent with solution 
\begin{equation}
C=C_0 \exp \left[ - \frac{t}{\tau_{lc}} \right],
\label{eq10}
\end{equation}

\begin{equation}
\tau_{lc} =  \frac{\rho_w U_w }{k^{\prime} \lambda_s (T-T_{lim})} H,
\label{eq11}
\end{equation}

\noindent and equation for high heat conductive \mbox{substrate $(\lambda_s \gg \lambda_w )$}

\begin{equation}
\frac{d C}{d t} = -  \frac{k^{\prime} \alpha}{\sqrt{h_0^2-\alpha \: t}} \: C,
\label{eq12}
\end{equation}

\noindent with solution
 
\begin{equation}
C=C_0 \exp \left[ - k^{\prime} h_0 \left( 1- \sqrt{1-\frac{\alpha}{h_0^2}t}\right)  \right],
\label{eq13}
\end{equation}

In the limit of short times, it follows from (\ref{eq13}) that

\begin{equation}
C \approx C_0 \exp \left[ - \frac{t}{\tau_{hc}} \right],
\label{eq14}
\end{equation}

\begin{equation}
\tau_{hc} = \frac{2 h_0 }{k^{\prime} \alpha} = \frac{2 h_0 \rho_w U_w}{k^{\prime} \lambda_w (T-T_{lim})}.
\label{eq15}
\end{equation}

Here  $\tau_{lc}$ and $\tau_{hc}$ are the characteristic lifetimes of viruses on a low heat-conducting and high heat-conducting substrate respectively.

\begin{figure}
\includegraphics[width=3.2 in,angle=0]{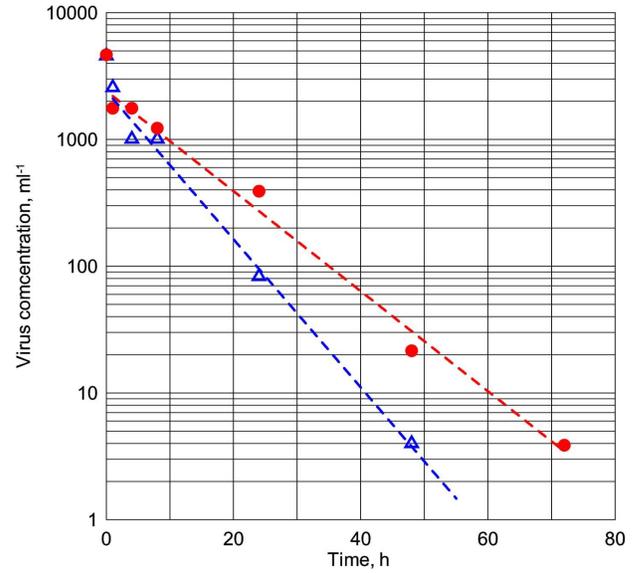}
\caption{Processing of experimental data \cite{COVID-1_Surf} to determine the decay law in the concentration of SARS-CoV-2 virus. 
$\triangle$: stainless steel substrate;  
$\circ$: plastic substrate. 
\label{Fig2}}
\end{figure}

%20
Additional treatment of experimental data \cite{COVID-1_Surf} shows an exponential decay of virus concentration over time (Fig. \ref{Fig2}).
Moreover, for a stainless steel substrate, this characteristic time is $\tau_{hc} \approx$ 7.4 h, for a plastic substrate  $\tau_{lc} \approx$ 11.0 h.

%21
For a low heat-conducting substrate, the proposed model also predicts a simple exponential decay of  the virus concentration  over  time. On a high conductive substrate, the concentration behavior is a more complex one. Asymptotically at initial times, virus concentration also has an exponential decay.
But from (11) it follows that in this problem there is one more characteristic time

\begin{equation}
\tau_{lim} = \frac{h_0^2}{\alpha} = \frac{\rho_w U_w h_0^2 }{\lambda_w (T-T_{lim})}.
\label{eq16}
\end{equation}

This time is equal to twice the time of liquid film evaporation in the considered approximation.

%22
According to our model after time $\tau_{lim}$  the concentration of the virus become constant. 
To note that similar picture is observed in the experimental data.

% % % % % % % % % % % % % % % % % % % % % % % % % % % % % % % %
%\begin{turnpage}
%\end{turnpage}
% % % % % % % % % % % % % % % % % % % % % % % % % % % % % % % % 

%23
The obtained expressions allow us to draw several interesting conclusions. 
The large viral lifetimes observed in the discussed experiments are consistent with formulas (\ref{eq11}) and (\ref{eq15}) under the condition of high humidity in the local environment of the substrate. 
An increase in humidity will lead to an increase in the lifetime of the viral population in both cases. 
It follows from the expression (\ref{eq15}) that the discussed lifetime for case of a high conductive substrate does not depend on the thickness of the substrate, but depends on the initial thickness of the liquid film. 
For a low heat-conducting substrate, it follows from the Exp. (\ref{eq11}) that the lifetime of viruses in liquid sample increase with a decrease in the thermal conductivity of the substrate and an increase in the total thickness of the substrate.

%24
These conclusions require additional verification. 
Nevertheless they can play a important role for medical practitioners and researches.

Finally let us estimate the rate of evaporation of aerosol, which can be formed on various surfaces by sneezing or coughing. 

%25
Consider a practical example when, in isothermal conditions, an "aqueous" aerosol with a coronovirus is deposited on a solid surface. 
Using again the approach related to the Stefan problem again \cite{Ref11_Model}, we obtain the estimation for the aerosol evaporation time $\tau_a$

\begin{equation}
\tau_{a} = \frac{\rho_w U_w H R }{2 \lambda_s (T-T_{lim})}.
\label{eq17}
\end{equation}

\begin{figure}
\includegraphics[width=3.2 in,angle=0]{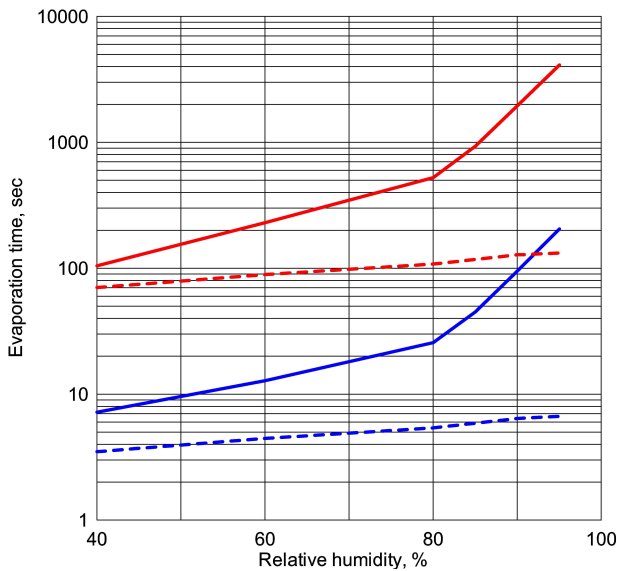}
\caption{Aerosol evaporation time on a 2 mm thick plastic substrate, depending on air humidity. 
Blue color is for 5 $\mu$m aerosol droplets, red color is for 100 $\mu$m droplets. 
Solid lines are for substrate and air temperature 22 $^{\circ} C$; 
dotted lines are for substrate temperature 37 $^{\circ} C$, air temperature 22 $^{\circ} C$. 
\label{Fig3}}
\end{figure}

As it can be seen, the thermal conductivity of the substrate, the initial radius of the aerosol droplets $R$, and the thickness of the substrate determine the time of the aerosol evaporation. 
During aerosol evaporation deformations near the evaporation front destroy viruses.  
It is known that when coughing, the average droplet size of the aerosol is 5 $\mu$m \cite{Ref16_Droplets}, while sneezing the aerosol has a generally bimodal distribution of drops with maxima in the size range of 100 and 600 $\mu$m \cite{Ref17_Droplets}. 
In Figure \ref{Fig3} for two temperatures 22 $^{\circ} C$ and 37 $^{\circ} C$ the evaporation time of water aerosol particles with a diameter of 5 and 100 $\mu$m are shown. 
The plastic substrate in this estimations has 2 mm thickness. Interestingly that the human skin has a thermal properties very similar to the plastic material \cite{Ref18_Skin}.
Therefore, these calculations will be relevant for estimation the aerosol evaporation time on the surface of human skin. 

%\begin{turnpage}
%\end{turnpage}

%26 
The evaporation time range varies widely and is sensitive to air humidity near the substrate. At 100\% relative humidity, the aerosol does not evaporate at all.

%27
The presented theory cannot cover all physical aspects of the complex problem of the vital functions of viruses. 
But we believe that it revealed an important correlation between the rate of evaporation of liquid biological sample and the vitality of coronavirus. 
It was shown that the main experimental results on the dynamics of coronavirus survival on various surfaces, presented in \cite{COVID-1_Surf}, correlate well with the rate of isothermal evaporation of liquid samples with viruses on substrates which do not absorb moisture. 
The evaporation rate depends on the thermal conductivity and thickness of the substrate, as well as the relative humidity of the surrounding air. 
Ceteris paribus, the viral culture should die faster on more thermally conductive and thinner substrates, as well as in lower relative humidity environment, which provide a higher evaporation rate.

\begin{acknowledgments}
The authors are grateful to Professor Oleg Penyazkov for support of this work.
\end{acknowledgments}

%\bibliography{Evaporation_Ref}
{}

%\newpage

% % % % % % % % % % % % % % % % % % % % % % % % % % % % % % % % % % % % % %
%\begin{turnpage}
%\begin{figure}
%\includegraphics{Fig1.eps}
%\caption{The sketch for evaporation of a liquid sample with a viral culture from the solid surface. 
%\label{Fig1}}
%\end{figure}
%\end{turnpage}
% % % % % % % % % % % % % % % % % % % % % % % % % % % % % % % % % % % % % % 

\end{document}